\documentclass[reprint,fleqn,superscriptaddress,twocolumn,showpacs,
amsmath,amssymb,prb,aps,short bibliography]{revtex4-2}
\usepackage[english]{babel}
\usepackage[all]{xy}
\usepackage{graphicx,psfrag,times,epsfig,color}
\usepackage{verbatim,natbib}
\usepackage{booktabs}
\usepackage{color}
\usepackage{makeidx}
\usepackage{amsmath}
\usepackage{bm}
\usepackage{amsfonts}
\usepackage{amssymb}
\usepackage{hyperref}
\usepackage{bm}
\usepackage{ragged2e}

\begin{document}

\let\namerefOld\nameref
\renewcommand{\nameref}[1]{\textit{\namerefOld{#1}}}

\title{A unified model for lunderdoped and overdoped cuprate superconductors based on a spinodal transition}
\author{H\'ercules H. Santana and E. V. L. de Mello}
\affiliation{Instituto de F\'{\i}sica, Universidade Federal Fluminense, 24210-346 Niter\'oi, RJ, Brazil}

\email[Corresponding author: ]{evlmello@id.uff.br}

\begin{abstract}

Many years of intense research on cuprate superconductors have led
to several discoveries, such as the pseudogap and charge
density waves (CDW), yet a complete theory is still lacking.
By analyzing some experiments and performing calculations, we provide a full
interpretation of their properties; from the undoped insulator to the overdoped metallic compounds.
The variation of the anomalous Hall coefficient ($R_{\rm H}(T)$) with temperature
at half-filling ($n = 1$) and, combinations of undoped ($p = 0$)
insulators and metallic films, which, among other things,
are indicative of a thermodynamic transition. On the overdoped side,
recent experiments near the superconducting-to-metal transition detecting
superconducting puddles and a considerable degree of charge disorder,
suggest that a similar thermodynamic transition
operates at all doping levels. We propose a spinodal or charge-separation transition starting
near the pseudogap temperature $T^*(p)$, which among other things generates
the CDW domains with a typical double-well Landau free-energy functional.
Thus, from  the half-filled to the overdoped region,
the free energy forms an array of wells with $n = 1$
{\it static} holes. With doping, {\it mobile} holes tend to occupy
these wells with alternating high and low densities, generating the CDW pattern.
The confined holes in small regions develop local superconducting amplitudes,
giving rise to a mesoscopic granular superconductor. Similar to the XY model, the grains
develop correlation effects mediated by Josephson coupling, which is proportional to
the local superfluid density.
This approach yields a unified theory of cuprate superconductors.

\end{abstract}
\pacs{}
\maketitle

\section{Introduction}

Cuprate superconductors are formed by doping holes or electrons
into a half-filled ($n =1$) parent insulator compound, and both groups of
materials display superconductivity, charge ordering and
antiferromagnetism\cite{Keimer2015}. For hole-doped cuprates,
there is enough experimental evidence showing that charge inhomogeneities
are present in the very low doping Mott insulator region\cite{CO.insulator2016,Insul.Stripes2019,Kang2023B}
that is a precursor of the insulator-superconductor transition, as well
as in the overdoped and heavily overdoped regions\cite{OverJJ2018,OverJJ2022,Over2023}.
In other words, the superconducting phase is intertwined or ``sandwiched''
between regions of charge inhomogeneities, and we argue here that it is the consequence
of a spinodal decomposition transition\cite{Bray1994}. Such a transition consists
of a single uniform phase at sufficiently high temperatures, and decomposes into
a charge-separated or two-phase component when quenched.

We begin our discussion by considering the Hall
coefficient ($R_{\rm H}$) measurements\cite{Hall.2007} in the prototype La$_{2-x}$Sr$_x$CuO$_4$ (LSCO),
where $x = p$ is the hole-doped
level. They studied several compounds with $0.0 \le p \le 0.21$
and measured the largest carrier variation with the temperature in the half-filled
($p = 0$ and $n = 1$ holes per Cu atom) and the $p = 0.01$ compounds\cite{Hall.2007}.
Notice that these two systems are insulators and, in principle, there are
no free charges. We propose here that a thermodynamic
spinodal decomposition\cite{Bray1994}
transition occurs at half-filling with pronounced charge fluctuations.
Such transition creates free energy modulations below the onset temperature,
taken to be the pseudogap $T^*(p)$ extrapolated to $p = 0$. The process
of doping with Sr creates compounds with $n = 1 + p$ holes per Cu atom where
the additional $p$ holes are mobile, giving rise to the ubiquitous charge density waves(CDWs)\cite{Comin2016}.
The two distinct holes are detected in many
experiments\cite{Kato2008,Huefner2008,Intimate2019}, and both types of holes may become indistinguishable
with the same energy near $p = 0.19$\cite{Anjos2025,Shen2019,Tranquada2024}.

Additional evidence that a spinodal decomposition sets in for half-filling systems is
provided by the realization of a superconducting transition by combining insulator
($p = 0$) LSCO films (I) with metallic (M)
films with $p = 0.45$, both non-superconductors when alone\cite{Gozar2008}. 
The composed systems are made by atomically
precise layer-depositing, and films of various widths, in any order of deposition
exhibit superconductor properties\cite{Gozar2008}. The explanation is that the
LSCO undoped half-filled insulator (the low density component) has
multiples empty-well structure like those of Fig. \ref{fig1}, while $La_{1.55}Sr_{0.45}CuO_4$ (the high density component)
has metallic free holes that fill the wells, resulting in a finite-$p$ interface layer\cite{Smadici2009}.

We have previously studied electronic phase separation using the Cahn-Hilliard diffusion equation,
which is the standard method to study spinodal decomposition transitions.
This method was used to study the CDW instabilities,
and we have shown that pairing interactions between spatially confined holes are
proportional to the ``degree'' of confinement, given by the average free-energy potential
$\left< V_{\rm GL} (p)\right>$\cite{TimeEvol2012,Mello2017,Mello2020a} (see Fig. \ref{fig1}).
This is based on earlier observations that smaller atomic grains of Sn have a higher critical field
that increases with decreasing size\cite{SmallGrains1969}. Similarly,
small atomic Bi clusters, which in contrast to non-supercoducting bulk, develop
superconductivity, which may also be a realization of topological superconductors\cite{BiGrains1991}.

Therefore, we recognize the rise of local superconducting planar amplitudes
$\Delta_{\rm sc}({\bf r}_i)$ in the mesoscopic
CDW/CO charge domains as a crucial ingredient of cuprate superconductivity.
To calculate $\Delta_{\rm sc}({\bf r}_i)$ we use a self-consistent
Bogoliubov-deGennes (BdG) method with the pairing interaction
proportional to $\left< V_{\rm GL} (p)\right>$, as explained above. We also use
a fixed local chemical potential to each system\cite{Mello2017,Mello2020a}, which in general, reproduce
the observed\cite{Comin2016} CDW/CO wave length $\lambda_{\rm CDW}(p)$.
The next step is to calculate the critical temperature $T_{\rm c}$, which is done following
the phase fluctuations of the local superconducting order parameters
described by an effective XY Hamiltonian\cite{Spivak1991}.
This is a Josephson coupling $J_{\rm i j}(\theta_i -\theta_j)$ between grain $i$ and $j$, where
$\theta_i$ is the local superconducting phase and $J_{\rm i j}$ is the
lattice version of the local superfluid density\cite{Spivak1991}.
This approach was used before to
reproduce the well-known proportionality between superfluid density\cite{Mello2021} and $T_{\rm c}$,
which is known as Uemura plot\cite{Uemura1989}, and the Bo{\v{z}}ovi{\'c}'s relation for overdoped compounds\cite{Bozovic2016}.

Here we are mainly concern with how superconductivity disappears at large hole doping,
which was systematically studied by Ref. \onlinecite{OverJJ2022},
combining magnetic susceptibility,
neutron difraction, angle-resolved photoelectron spectroscopy (ARPES)
and spectroscopic imaging scanning tunnelling (SI-STM). In particular, measurements of bulk susceptibility on LSCO
crystals with $p = 0.25$ indicate an onset of diamagnetism at $T_{\rm c1} = 38.5$ K,
with a sharp transition to a phase with full bulk shielding at $T_{\rm c2} = 18$ K.
On highly overdoped LSCO films of $p = 0.35$ they found superconducting puddles and
differential conductances with large amplitudes\cite{OverJJ2022}.
Along the same lines, scanning tunnelling spectroscopy (STS)\cite{Over2023}
detected puddle formation and persistent gaps across the end of the dome,
concluding that the superconducting phase does not break-down by
the vanishing of the superconducting gap.
If we follow Ref. \onlinecite{Hall.1994} and take the pseudogap $T^*(p) \approx 100$ K,
in the interval $0.25 \le p \le 0.35$, it implies that a reminiscent of the
spinodal transition and charge fluctuations extend to $p \approx 0.35$. Furthermore,
$\left< V_{\rm GL} (p)\right>$ that scales with $T^*(p)$ remains constant in this interval,
whaich reproduces the onset temperature of the magnetic susceptibilities mentioned above\cite{OverJJ2022},
and the gap filling detected by Ref. \onlinecite{Over2023}.

We recall that the differences in underdoped and overdoped
properties of cuprates like LSCO makes any general theory very difficult.
About thirty years ago, Emery and Kivelson proposed a successful model based on 
very strong phase fluctuations that, they argued, were consequence of 
these materials being doped insulators with very small superfluid density\cite{Emery1995}.
While the model explained several features of underdoped compounds, the 
predictions for the overdoped properties did not agree with the recent
experiments\cite{OverJJ2022,Over2023}. A more recent 
theoretical approach for highly overdoped cuprates\cite{Spivak2008}
proposed that superconductivity initially develops 
within disconnected self-organized nanoscale grains, and long-range order
is achieved through the proximity effect. However, this model is just
applicable to overdoped materials.

In this paper, we argue that properties of either low- or high-doping cuprates are described by
the same phase separation or spinodal transition that sets in at half-filling and
propagates or remains into large overdoped materials, even beyond the superconducting dome.
The calculations constitute a general model for cuprates, covering
all superconducting phase diagram. To illustrate the method, we present here calculations on the recent
experiments at the end of the superconducting phase\cite{OverJJ2022,Over2023}.

\section{The Calculations }

\begin{figure}[!ht] 
\centerline{\includegraphics[height=4.5cm]{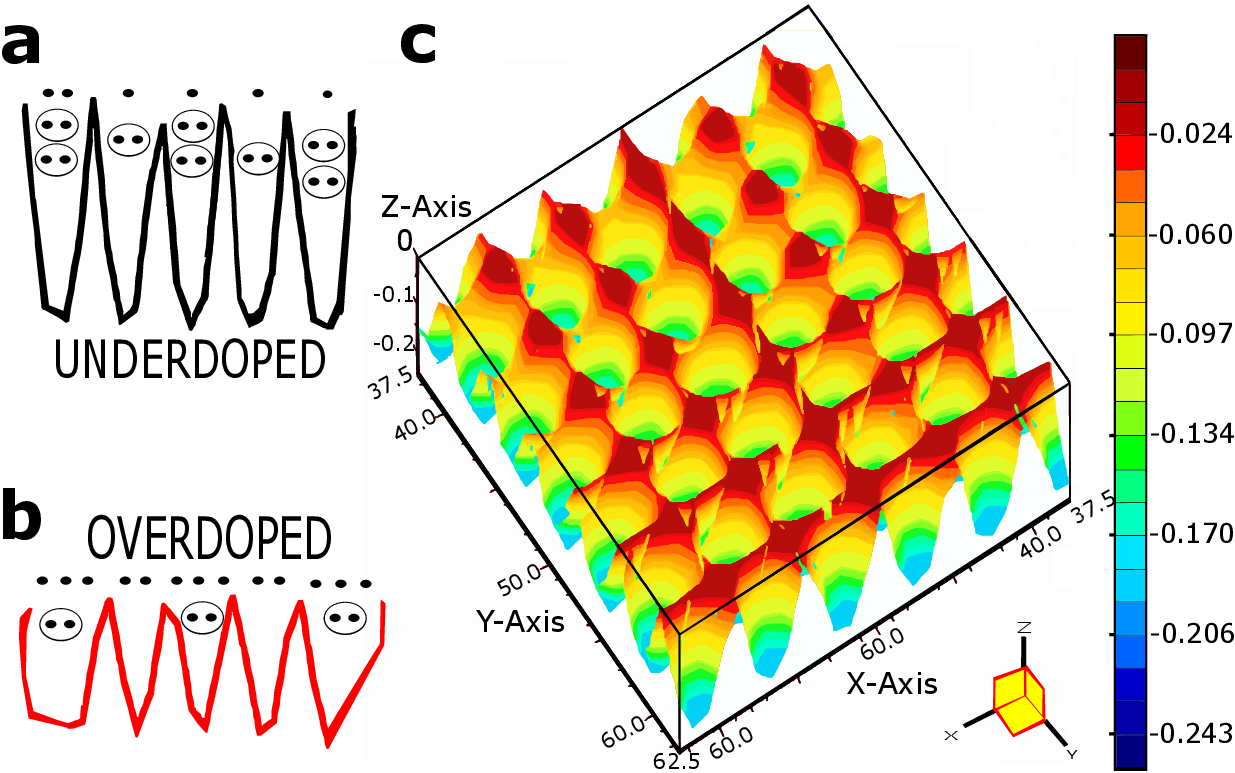}}
\caption{
{\bf The CH simulations of the phase separation GL free energy
potential}. The $V_{\rm GL}(u({\bf r}))$, that leads to the bimodal distribution.
We show in c) a central portion of a $N \times N$ array of unit cells that are square-shaped. In a) and b)
$V_{\rm GL}(r_i)$ along the $x$ direction of the CuO plane.
These insets show the tendency to form pair-density waves (PDW)\cite{Mello2021} at low temperature.
}
\label{fig1}
\end{figure}

The spinodal decomposition transition is generally simulated by solving the Cahn-Hilliard (CH)
nonlinear differential equation. Here, we use
a stable and fast conservative ﬁnite-difference scheme\cite{Otton2005}.
The free energy potential $V_{\rm GL}$ develops symmetric modulations similar to
those shown in Fig. \ref{fig1}, and it is possible to reproduce other
symmetries such as unidirectional stripes\cite{Mello2017}. The resulting bimodal
distribution separates the rich and poor phases in domains or grains, bound
by the free energy wells like those of Fig.\ref{fig1}, leading
to the CDWs with wavelength $\lambda_{\rm CDW}(p)$.
In this approach, the pseudogap temperature $T^*(p)$ is the transition temperature,
acting as the boundary between uniform charge density and the bimodal decomposition.

In previous works\cite{DeMello2012,Mello2017,Mello2020a},
we described the CH method in detail
and its connection with the cuprates' superconducting properties.
Briefly, just for completeness, the starting
point is the time-dependent phase separation order parameter associated
with the local electronic density, $u({\bf r}) = (p({\bf r}) - p)/p$, where $p({\bf r})$
is the local charge or hole density at a position ${\bf r}$ in the CuO plane and
$p$ is the average doping $\left < p \right >$.
The CH equation is based on the minimization of the Ginzburg–Landau
(GL) free-energy expansion in terms of
the conserved order parameter\cite{TimeEvol2012} $u(r, t)$;

\begin{equation}
f(u)= {{\frac{1}{2}\varepsilon |\nabla u|^2 +V_{\rm GL}(u,T)}},
\label{FE}
\end{equation}
where  ${V_{\rm GL}}(u,T)= -A(T) u^2/2 + B^2u^4/4+...$ and $A(T) = \alpha (T_{\rm PS}-T)$, is a
double-well potential that characterizes the electronic phase separation into
hole-rich and hole-poor phases
below $T_{\rm PS}$, assumed to be $T^*(p)$. We use
$\alpha$ and $B =1$, and $\varepsilon$ controls the spatial separation of the
charge-segregated patches that are connected with the experimental\cite{Comin2016}
wavelength $\lambda_{\rm CO}(p)$
and neglect the temperature dependence\cite{TimeEvol2012,Mello2017,Mello2020a}.

In Fig. \ref{fig1}, we show the 3D simulation of the spinodal or
CH phase-separation potential $V_{\rm GL}(u({\bf r}))$ that generates the CDW/CO.
Along one direction, for instance, the $x$-direction,
we can see how the high and low $V_{\rm GL}$ modulations confine alternating hole-rich and hole-poor
domains, as shown in the {\bf a} and {\bf b} insets.
We have previously argued\cite{TimeEvol2012,deMelloKasal2012}
that this highly inhomogeneous charge unbalance, favors pairing attraction
inside the charge domains. This interaction is
proportional to the mean $V_{\rm GL}(u({\bf r}))$ amplitude of oscillations $\left < V_{\rm GL} \right >$.
Typically $\left < V_{\rm GL} \right >$ is proportional to
the $T^*(p)$, and decreases with increasing doping\cite{TimeEvol2012,Mello2017,Mello2020a}.

From the above reasoning, we use the BdG method with a nearest-neighbour attractive Hubbard
potential given by\cite{Mello2020a} $V (p, T) = V_0 \times [1- T / T^*]^2$,
where $V_0 \propto \left< V_{\rm GL} \right >$. This is done self-consistently, with
parameters derived for overdoped LSCO from Ref. \onlinecite{Zhong2022}.
According to resonant x-ray scattering (REXS) experiments\cite{Comin2016}, the CDW/CO patterns do not
change appreciably with the temperature near and below the superconducting $T_{\rm c}$, therefore,
our BdG method consists in keeping a given CO structure constant by adjusting
the local chemical potential during each self-consistent
interaction while $p ({\bf r}_i)$ and $\Delta_{\rm sc}({\bf r}_i)$ are converging\cite{Mello2020a}.
Typical $p ({\bf r}_i)$ and $\Delta_{\rm sc}({\bf r}_i)$ maps are respectively
shown in Fig. \ref{fig025}.

Fig. \ref{fig025} also shows that $\Delta_d({\bf r}_i,p, T)$ have local spatial 
variations and, because the cuprates` short superconducting coherence length, each
region with an average amplitude can be regarded as an ``isolated'' grain. Therefore, as in granular
superconductors, the grains are coupled by Josephson energy between the alternating
charge regions with $E_{\rm J}(r_{\rm i j})$, which is the lattice analogue
of the local superfluid density\cite{Spivak1991} $\rho_{ sc} (r{\rm i j})$.
Under these considerations, the global superconducting properties such as
the condensation energies\cite{Mello2022}, critical temperatures\cite{Mello2021}, and interlayer Josephson
coupling are function of the average SC amplitude
$\left <\Delta_d(p,T)\right > = \sum_{i}^{N} \Delta_d (r_i, p, T)/N $, where $i$ runs
over $N$ unit cells of the CuO plane. 
Based on the work of Bruder {\it et al}\cite{Bruder95}, even for $d$-wave amplitudes, 
it is sufficient to use the Ambegaokar-Baratoff analytical $s$-wave expression\cite{AB1963}
averaged over the plane:
\begin{equation}
 {\left < E_{\rm J}(p,T) \right >} = \frac{\pi \hbar {\left <\Delta_d(p,T)\right >}}
 {4 e^2 R_{\rm n}(p)} 
 {\rm tanh} \bigl [\frac{\left <\Delta_d(p,T)\right >}{2k_{\rm B}T} \bigr ] .
\label{EJ} 
\end{equation}
Where $R_{\rm n}(p, \sim T_{\rm c})$ 
is proportional to the corresponding normal state resistance just above $T_{\rm c}$. 
Notice that ${\left < E_{\rm J}(p,T) \right >}$ is a global property and may be 
nonzero even when 
$ \Delta_d (r_i, p, T)$ vanish in some regions, provided that the averages 
$\left <\Delta_d(p,T)\right >$ is
finite. By the same token, ${\left < E_{\rm J}(p,T) \right >}$ can be vanishing small with a dilute
superconducting amplitude occupying a small volume fraction, which we believe 
is the situation of
the strongly overdoped compounds of Refs. \onlinecite{OverJJ2022} and  \onlinecite{Over2023}.

\section{Results}

\begin{figure}[!ht] 
\centerline{\includegraphics[height=3.70cm]{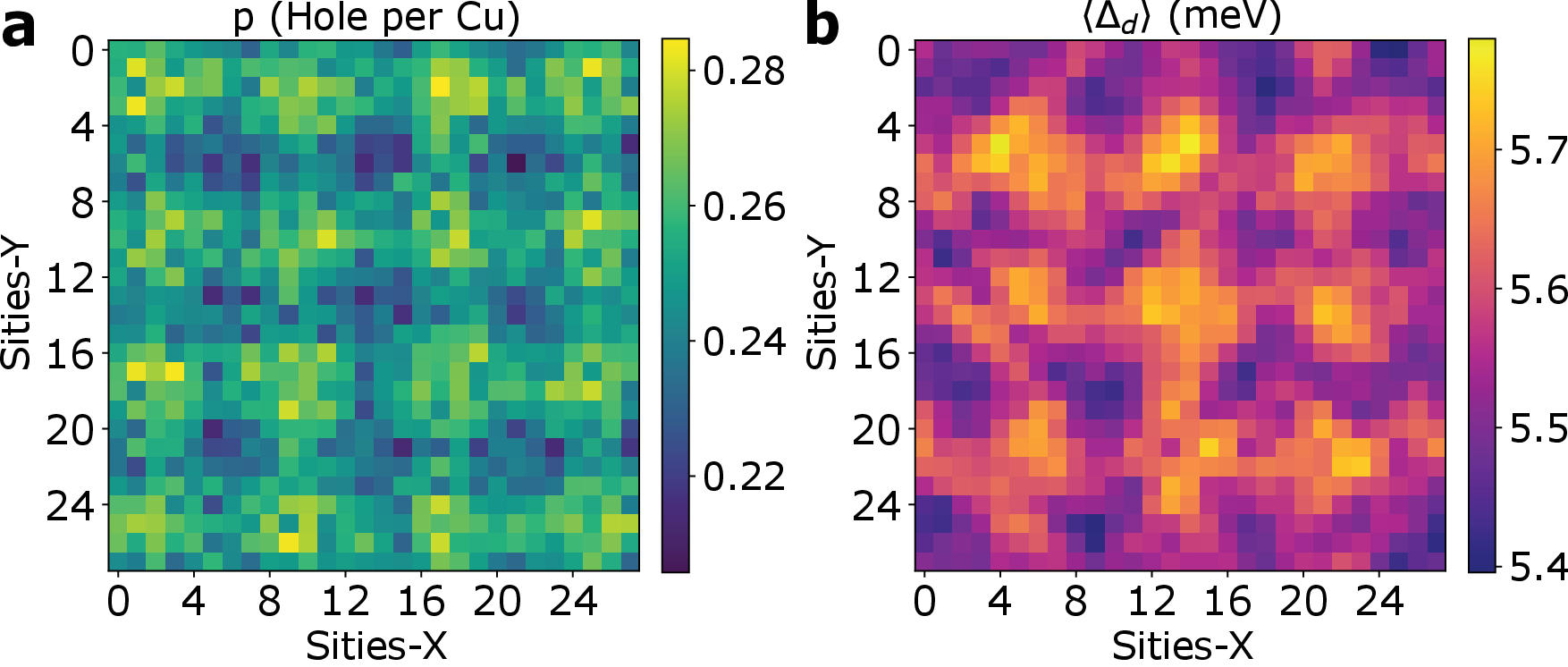}}
\caption { {\bf The local charge density and the superconducting amplitude.}
a) Charge distribution $p(r_i)$  of a compound with average
doping $p = 0.25$. b) The two-dimensional BdG $\Delta_d(p=0.25,0)$ calculations on the same
location. It demonstrates that, everything being the same, regions with local higher densities
have smaller local superconducting
amplitudes, and this deminstrate the suppression of the superconducting phase in strongly overdoped regions.
The results in {\bf b}, with gaps everywhere, is consistent with the break down
of superconductivity by gap filling\cite{Over2023}.
}
\label{fig025}
\end{figure}

Now, we apply these calculations to compounds near the superconducting-to-metal transition
experiments mentioned in the introduction.
We start with the measurements of Ref. \onlinecite{OverJJ2022} on
a LSCO $\left < p \right > = p = 0.25$ compound and simulate the charge density oscillations
which are shown in Fig. \ref{fig025}{\bf a}. On this charge distribution, we calculate the BdG superconducting amplitude map
$ \Delta_d (r, T=0)$ in Fig. \ref{fig025}{\bf b}. The calculations
with the average $\left < V_{\rm GL}(p)\right >$ around $\sim 5.6$ meV are in
agreement with the low-temperature STM measurements\cite{Kato2008}.

\begin{figure}[!ht] 
\centerline{\includegraphics[height=5.0cm]{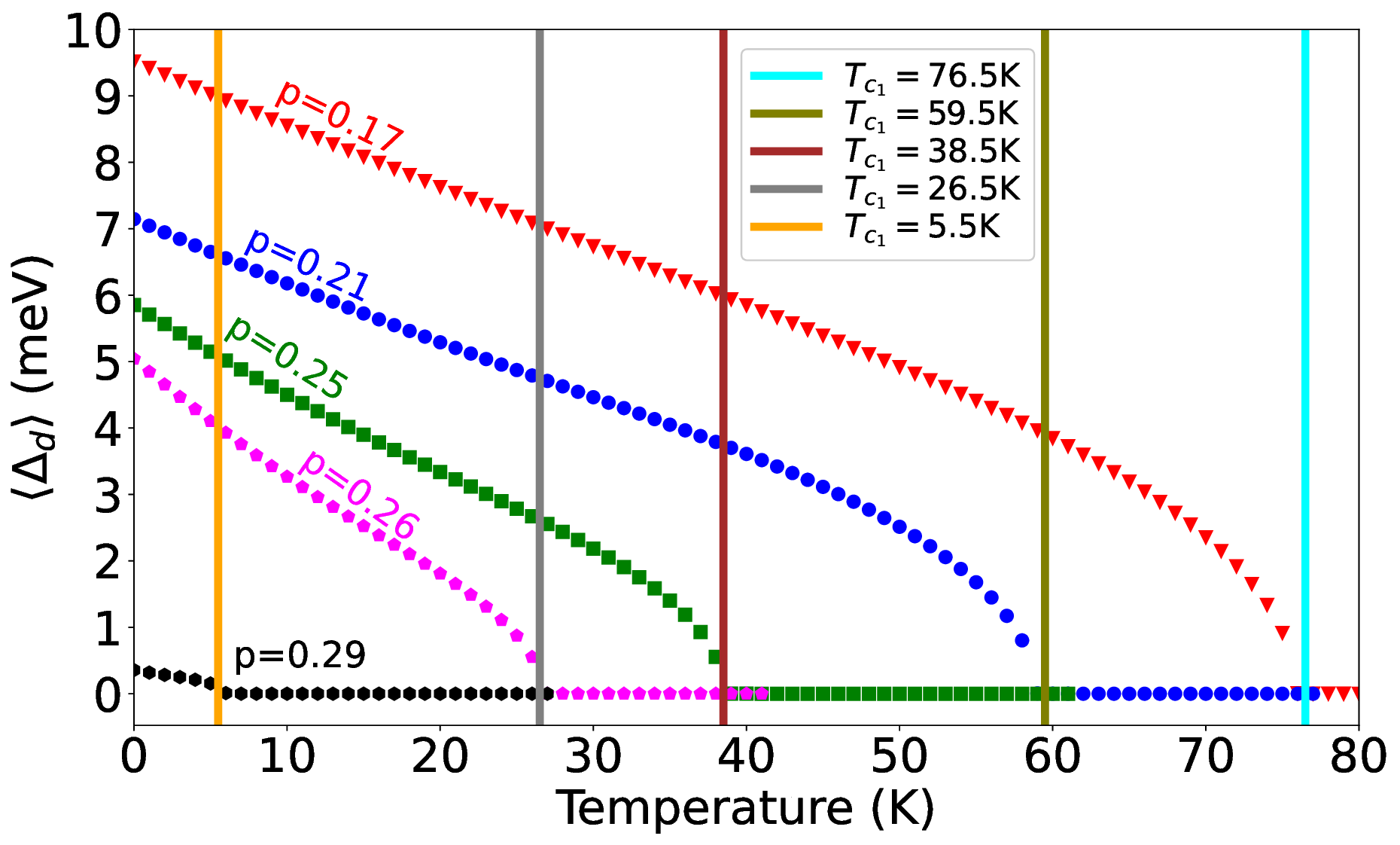}}
\caption{ {\bf The average $d$-wave superconducting amplitude for selected compounds}
$\left <\Delta_{\rm sc}(p, T)\right>$ as a function of temperature
for some overdoped compounds. The zero temperature results match the
experimental values\cite{Kato2008} and we derive the onset of superconducting fluctuations $ T_{\rm c}^{on}(p)$
when $\left <\Delta_{\rm sc}(p, T)\right> \rightarrow 0$ that coincides with the
onset of susceptibility signal of Ref. \onlinecite{OverJJ2022} and the Nernst
effect measurements\cite{Nernst2010,Rourke2011,NernstPRB2018}.}
\label{figAllD}
\end{figure}

\begin{figure}[!ht] 
\centerline{\includegraphics[height=5.0cm]{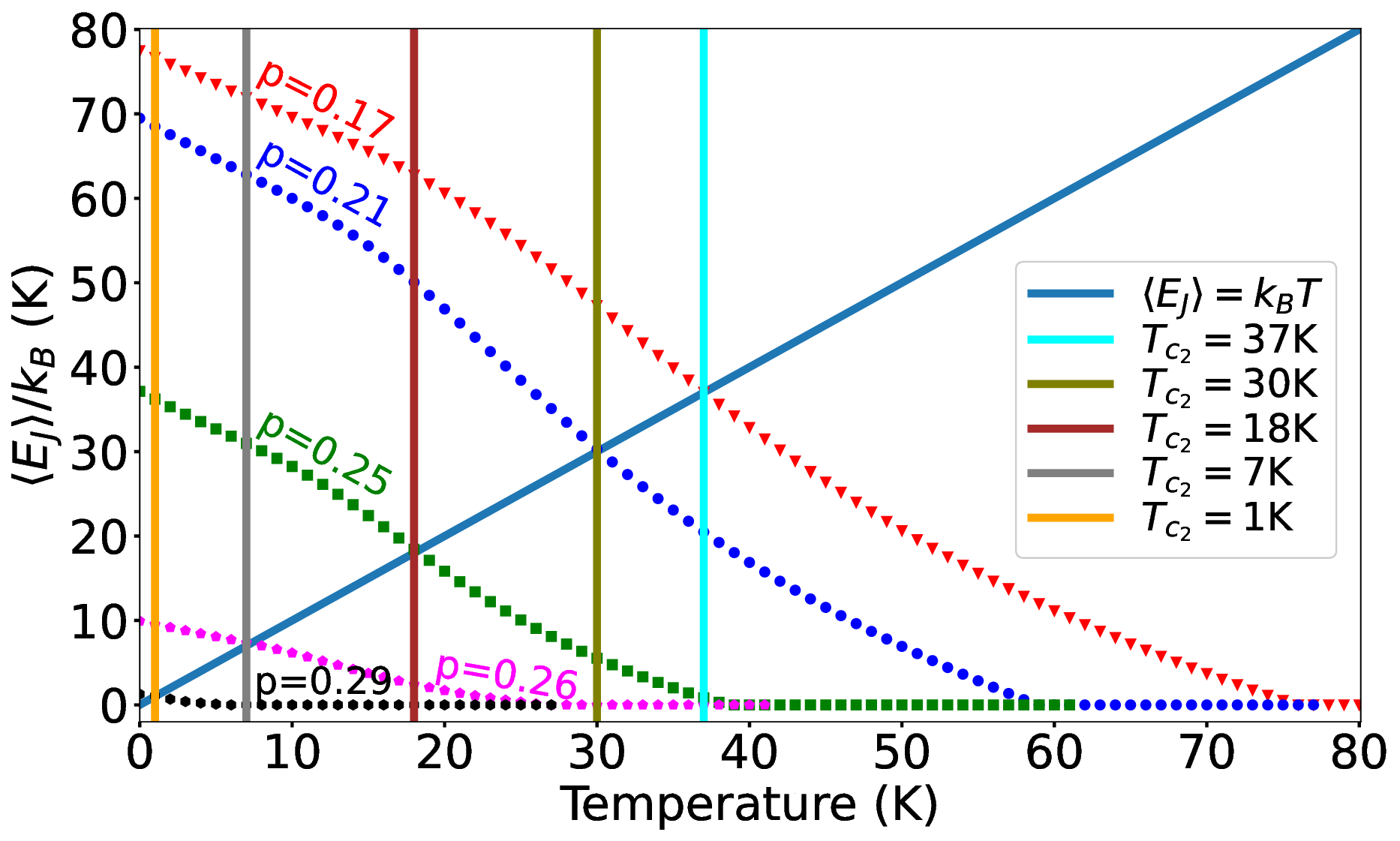}}
\caption{ {\bf The average Josephson coupling as a function of temperature.}
$\left < E_{\rm J}(p,T_{\rm c}) \right > $ curves
and the thermal fluctuation energy $k_{\rm B}T$. The superconducting
long-range phase-ordering transition at $ T_{\rm c}(p)$ occurs when the straight line $k_{\rm B}T$ crosses
the $\left < E_{\rm J}(p,T_{\rm c}) \right > $ curves.
}
\label{figAllEJ}
\end{figure}

Next, we perform finite temperature calculations and plot $\left <\Delta_d(p,T)\right > \times T$
in Fig. \ref{figAllD}, which with Eq. \ref{EJ} are used to
derive the critical temperatures. We recall that
$T_{\rm c}(p)$ is the superconducting long-range phase-ordering temperature, and it is obtained
by the competition between the
average Josephson coupling of the local phases $\theta_i$ in the charge domains
and the disorder arising from the thermal energy $k_{\rm B}T$.
By plotting both together in Fig. \ref{figAllEJ}, we derive $T_{\rm c}(0.25) \approx 19$ K  when they are equal,
i.e., ${\left < E_{\rm J}(p,T_{\rm c}) \right >} = k_{\rm B}T_{\rm c}$ by comparing both
sides of Eq. \ref{EJ}, which is plotted in Fig. \ref{figAllEJ}

In Fig. \ref{figAllD} we show that the average $\left <\Delta_d(p,T)\right >$  used above to
calculate $T_{\rm c}$ vanishes
at $T \approx 40$ K, well above $T_{\rm c}$ but in close
agreement with the measured diamagnetic susceptibility onset\cite{OverJJ2022}. Another interesting
result from these BdG calculations is that, everything being the same, higher local densities yield lower
local amplitudes, which can be seen comparing Fig. \ref{fig025}{\bf a} and Fig. \ref{fig025}{\bf b}.
Furthermore, the superconducting
amplitudes are present everywhere at low temperatures, in agreement
with other theoretical predictions for cuprates\cite{Spivak2008}. We emphasize that these findings and the
calculated gaps $ \Delta_d (r, T=0)$ shown in Fig. \ref{fig025}{\bf b} are also in agreement with the STS
series of experiments that demonstrate
that the superconductivity breaks down by gap filling\cite{Over2023}.

We repeat this procedure for the overdoped compounds
$\left < p \right > = p = 0.17, 0.21, 0.26$, and 0.29, starting always with
the zero-temperature $\left <\Delta_{\rm sc}(p, 0)\right >$
in good agreement with the STM results\cite{Kato2008}.
After this, the temperature-dependent $\left <\Delta_{\rm sc}(p, T)\right >$
are evaluated using $V (p, T)$ mentioned
above, and with $T^*(p)$ from the literature, and the results are
plotted in Fig. \ref{figAllD}. The temperature where $\left <\Delta_{\rm sc}(p, T)\right >$ vanishes
determines $ T_{\rm c}^{on}(p)$,
the onset phase-fluctuation region above $ T_{\rm c}$
that was studied by several methods, such as STM\cite{Gomes2007} and $\mu$SR\cite{Muon2013}
on Bi-based cuprates, and susceptibility\cite{OverJJ2018,OverJJ2022} and Nernst
effect\cite{Nernst2010,Rourke2011,NernstPRB2018} on LSCO.
The parameters used in the Hubbard model for the BdG calculation are taken from
Refs. \citenum{Hoek2016,Kato2005,Kato2007,Kim2004,Fujimori1998}.

The onset calculations yield good agreement with the measured\cite{OverJJ2018,OverJJ2022,Rourke2011}
phase-fluctuation region upon increasing the temperature above $T_{\rm c}(p)$
for $p = 0.17, 0.21, 0.25$, and 0.29 compounds, suggesting that the superconducting amplitude is
indeed nonzero above $T_{\rm c}(p)$ and below the $T^*(p)$. This may also occur at the end and beyond of superconducting
phase at $p_{\rm c} \approx 0.27$ due to spinodal transition fluctuations.

\section{Conclusions}

We develop a model for cuprates based on a spinodal decomposition or phase separation thermodynamic
transition below the temperature $T^*(p)$ that produces an array of free-energy wells, which leads to the
mesoscopic CDW. The superconducting pair-potential is proportional to the free-energy aplitude of the
charge domains or the spinodal barrier between the rich and poor density phases. The
superconducting BdG calculations yield the local gaps $ \Delta_d (r_i, T=0)$ which
are connected by Josephson coupling between the mesoscopic grains.
The critical temperature is derived by the competition between the thermal energy
$k_{\rm B}T_{\rm c}(p)$ and the average Josephson coupling. When the system
is quenched from a homogeneous phase above $T^*(p)$ in the overdoped region, where
the holes are mobile, large charge fluctuations occur\cite{Bray1994}, in agreement with
recent experiments that detected mesoscopic superconducting puddles 
even beyond the superconducting dome\cite{OverJJ2022,Over2023}.
These observations in the overdoped thin LSCO films\cite{OverJJ2022} are endorsed in the frame of
our model by the measurements of finite $T^*(p)$ up to $p \approx 0.35$\cite{Hall.1994}

Similarly, the local $\left <\Delta_{\rm sc}(p, T)\right>$ calculations shown in
Fig. \ref{figAllD} demonstrate that
the average superconducting amplitude vanishes at temperatures much higher than $T_{\rm c}(p)$,
providing a mechanism for the susceptibilities measurements of Ref. \onlinecite{OverJJ2022}.
The calculations presented in Fig. \ref{fig025}{\bf b} show the ubiquitous presence of $\Delta_{\rm sc}(r_i)$,
with smaller values at higher hole densities, in agreement
with the breakdown of the superconductivity by gap filling reported by Ref. \onlinecite{Over2023}.

To complement our proposal of a mesoscopic phase separation or spinodal transition,
on the other side of the dome, verified by the large Hall coefficient $R_{\rm H}(p)$
variation with temperature\cite{Anjos2025}. Furthermore, the model
provides a way to explain how the reported superconductivity\cite{Gozar2008} in
bilayers consisting of insulators (La$_2$CuO$_4$) (the low-density component) and metallic overdoped
(La$_{1.55}$Sr$_{0.45}$CuO$_4$) (the high-density component).

Therefore, we have shown above that the phase separation model is applicable to experiments performed
in the overdoped region \cite{OverJJ2022,Over2023}, complementing
our previous calculations on the Hall effect for
finite doping\cite{Anjos2025} ($0 \le p\le 0.21$. Together, both calculations
indicate that the transition to the superconducting state occurs
by the same mechanism in the entire superconducting phase, unifying the
approach to underdoped and overdoped cuprates\cite{Mello2020a,Mello2023}.

\section{acknowledgements}

We acknowledge partial support from the Brazilian agencies CNPq and 
by Funda\c{c}\~ao Carlos Chagas Filho de Amparo
Pesquisa do Estado do Rio de Janeiro (FAPERJ), Project No.
E-26/211.270/2021.

\section{Methods}
All the calculations are explained here and in our previous published works.
\section{References}

%


\end{document}